\documentclass[10pt]{article}
\pdfoutput 1
\usepackage{graphicx}
\usepackage{amsmath}
\usepackage{amssymb}
\usepackage{bbm}
\usepackage[colorlinks=true]{hyperref}

\DeclareFontFamily{OT1}{pzc}{}
\DeclareFontShape{OT1}{pzc}{m}{it}{<-> s * [1.10] pzcmi7t}{}
\DeclareMathAlphabet{\mathpzc}{OT1}{pzc}{m}{it}

\title{Cartesian isotropic tensors revisited}
\author{Antonio O.\ Bouzas\thanks{email: 
    abouzas@cinvestav.mx.
    ORCID: 0000-0001-5493-4958.
  }  \\\small Departamento de F\'{\i}sica Aplicada, CINVESTAV-IPN \\[-4pt]\small
  Carretera Antigua a Progreso Km.\ 6, Apdo.\ Postal 73
  ``Cordemex''\\[-4pt]\small M\'erida 97310, Yucat\'an, M\'exico}
\date{3 September 2026}
\date{\today}

\begin{document}

\maketitle
\begin{abstract}
  An alternative, streamlined methodology is presented for deriving
  the isotropic Cartesian tensor bases under the special orthogonal
  group $\text{SO}(3)$. By shifting the traditional interpretation of
  the isotropy condition to analyze it as an algebraic system of
  equations for the rotation matrices themselves rather than the
  tensor components, lower-rank bases can be explicitly established in
  just a few lines without requiring finite coordinate rotations or
  complicated contractions of infinitesimal generators. Furthermore, a
  transparent combinatorial interpretation is provided for the number
  of tensors in the general higher-rank spanning sets. Finally, the
  Gram-matrix method is advocated as an efficient computational sieve
  to resolve the subsequent linear dependence.
  
\end{abstract}

\paragraph{Introduction.}\label{sec:intro}

The first textbook on the theory of Cartesian tensors of the
three-dimensional rotation group in flat Euclidean space, as currently
used by physicists, is that of Jeffreys~\cite{jeffreys_1931}. Published 
nearly a century ago, \cite{jeffreys_1931}~provides a derivation of the bases 
for isotropic tensors of ranks~2, 3, and, for the first time, 4, using a 
method based on finite rotations that is rather laborious.\footnote{This 
  treatment of isotropic tensors is generalized to dimension~$n>3$ and 
  arbitrary rank in~\cite{jeffreys_1973}.} In his classic Spanish-language
textbook~\cite{sant}, Santal\'o derives the bases for rank-2 and
rank-3 tensors using the same methods as in~\cite{jeffreys_1931},
although that reference is not explicitly cited. The same basis for
the space of rank-4 tensors given in~\cite{jeffreys_1931} is
reproduced in~\cite{sant} without a derivation; however, it is
mentioned that such a derivation involves a ``somewhat lengthy
reasoning'' which, it can be reasonably assumed, is the one provided
in~\cite{jeffreys_1931}.
 
In~\cite{hodge_1961}, an alternate derivation of the bases for
isotropic tensors of ranks~2, 3, and~4 is given, based on
infinitesimal rotations. That derivation is substantially shorter than
the one in~\cite{jeffreys_1931}, but it entails a significant degree
of algebraic complexity that is left mostly implicit. Consequently, it
remains unclear how applicable this approach would be to tensors of
even moderately higher ranks. More recently, the issue of the bases
for isotropic tensors of ranks~2, 3, and~4 has been addressed
in~\cite{rmfe} from an explicitly pedagogical perspective. The
approach followed in~\cite{rmfe} utilizes the infinitesimal rotations
of~\cite{hodge_1961}; however, these rotations are not imported
directly from the $\text{so}(3)$ Lie algebra but, rather, are
explicitly constructed from elementary vector calculus
considerations. Similarly, all other algebraic steps omitted in the
derivations of~\cite{hodge_1961} are fully provided, yielding a
pedagogically self-contained presentation. With its 15~pages and more
than 100~equations, the exposition in~\cite{rmfe} is certainly quite
lengthy, though this is a direct consequence of its educational focus.

In this paper, we point out that the derivation of bases for
low-rank isotropic tensors need not be lengthy or complicated. We
posit that the isotropy conditions must be viewed not as algebraic
equations for the isotropic tensors but, rather, as equations for the
$\text{SO}(3)$ rotation matrices themselves. In this way, we demonstrate 
below that a clear derivation of a basis for rank-4 isotropic tensors,
in particular, takes no more than about a dozen lines, and about twice
as many in the case of rank~5 due to the issue of linear
dependence. Demonstrating that these lower-rank bases can be
established so concisely provides significant pedagogical
clarity. Furthermore, by stripping away unnecessary algebraic
complications, the path toward more interesting research directions
becomes much clearer.

This paper is organized as follows. In the next section, we discuss the
definition and fundamental algebraic properties of the group
$\text{SO}(3)$ that are required in the rest of the paper. Our
discussion of isotropic tensors begins at rank~1 and proceeds by
increasing the rank successively by one unit, as is customary. We provide
explicit derivations of bases for isotropic tensors of ranks~1 to~5 in
Secs.~\ref{sec:rank.1}--\ref{sec:rank.5}. In Sec.~\ref{sec:rank.5}, 
we also discuss spanning sets and linear independence. We analyze
the algebraic structure of isotropic tensors of ranks~6--8 in 
Secs.~\ref{sec:rank.6}--\ref{sec:rank.8}. In those cases, however, we do not
provide the bases explicitly but rather refer the reader to~\cite{kea75}. 
In Secs.~\ref{sec:rank.7} and~\ref{sec:rank.8}, we also briefly discuss the 
construction of spanning sets for arbitrary odd and even ranks, respectively. 
Finally, in Sec.~\ref{sec:fin.rem}, we give a summary and our concluding remarks.

\paragraph{The group $\boldsymbol{\mathrm{SO}(3)}$.}\label{sec:so3}

In what follows, we restrict our attention to the group of
special orthogonal transformations, $\text{SO}(3)$. This is the group
of matrices $R\in\mathbb{R}^{3\times3}$ satisfying
\begin{equation}
  \label{eq:so3}
  RR^T=\mathbbm{1}, \qquad \det(R)=1,
\end{equation}
where $\mathbbm{1}$ is the $3\times3$ identity matrix. For our
purposes, we require Eq.~(\ref{eq:so3}) written explicitly in
tensor index notation. To this end, we recall the well-known
relation  (see, for instance, \cite{byron_1992} Eq.\ (3.18)),
\begin{equation}
  \label{eq:deta}
  \varepsilon_{j_1j_2j_3} A_{i_1j_1}A_{i_2j_2}A_{i_3j_3} = \det(A)
  \varepsilon_{i_1i_2i_3}, 
\end{equation}
valid for any matrix $A\in\mathbb{R}^{3\times3}$. By using
Eq.~(\ref{eq:deta}), we rewrite Eq.~(\ref{eq:so3}) as
\begin{equation}
  \label{eq:so3.ind}
  \text{(a)}\; \delta_{j_1j_2} R_{i_1j_1} R_{i_2j_2} = \delta_{i_1i_2},
  \qquad
  \text{(b)}\;   \varepsilon_{j_1j_2j_3} R_{i_1j_1}R_{i_2j_2}R_{i_3j_3} = 
  \varepsilon_{i_1i_2i_3},
  \qquad
  \forall R\in\text{SO}(3).  
\end{equation}
We point out here that, given $R\in\text{SO}(3)$, by definition
$R^T=R^{-1}\in\text{SO}(3)$ must also satisfy Eq.~(\ref{eq:so3.ind}), so
that we have
\begin{equation*}
  \label{eq:so3.ind.trans}
  \text{(a')}\; \delta_{i_1i_2} R_{i_1j_1} R_{i_2j_2} = \delta_{j_1j_2},
  \qquad
  \text{(b')}\;   \varepsilon_{i_1i_2i_3} R_{i_1j_1}R_{i_2j_2}R_{i_3j_3} = 
  \varepsilon_{j_1j_2j_3},
  \qquad
  \forall R\in\text{SO}(3), 
\end{equation*}
though we do not regard these equations as independent of
Eq.~(\ref{eq:so3.ind}).

We stress also that Eqs.~(\ref{eq:so3.ind}) not only serve as the
definition of $\text{SO}(3)$, but they also establish the isotropy of
the Kronecker and Levi-Civita tensors. We could, in fact, define the
$\text{SO}(3)$ group as the set of $3\times3$ real matrices having the
Kronecker and Levi-Civita tensors, and only those, as their fundamental 
isotropic tensors. Such a definition would be mathematically equivalent to the
standard Eq.~(\ref{eq:so3}), because both lead to the same system of
equations~(\ref{eq:so3.ind}). This provides the underlying rationale for why there is 
no other non-trivial fundamental isotropic tensor of $\text{SO}(3)$: if one
existed, its isotropy equation would lead to an additional defining
equation, (\ref{eq:so3.ind})(c), and, therefore, restrict the group to a
proper subset of $\text{SO}(3)$. (An explicit example of this is given at the
end of the next section.)

\paragraph{Isotropic vectors.}\label{sec:rank.1}

The existence of an isotropic vector $\vec{v}$ gives rise to
the equality,
\begin{equation}
  \label{eq:iso.rank.1}
  R_{ij} v_{j} = v_i,
  \qquad
  \forall R\in\text{SO}(3).
\end{equation}
Since this equation must be valid for all rotations $R$, we may
consider the rotations $R_x$, $R_y$, $R_z$ of 90$^\circ$ about the $x$,
$y$, $z$ axis, respectively. Equation (\ref{eq:iso.rank.1}) with
$R=R_x$ yields $v_y=0=v_z$, with $R=R_y$ yields $v_x=0=v_z$, and with
$R=R_z$ yields $v_x=0=v_y$, therefore $\vec{v}=0$. What we have done here
is to consider three particular values of $R$ in (\ref{eq:iso.rank.1})
to obtain a system of equations for the vector $\vec{v}$. This is,
essentially, the point of view adopted in
\cite{jeffreys_1931,jeffreys_1973} and, presumably, 
also in \cite{sant}. In this rank-1 case, the infinitesimal approach
of \cite{hodge_1961,rmfe} is essentially equivalent, since for a unit
vector $\hat{\phi}$ orthogonal to $\vec{v}$, an infinitesimal rotation
leads to $\delta v_i/\delta \phi = -\varepsilon_{ijk} \hat{\phi}_j
v_k$, which is a 90$^\circ$ rotation of $\vec{v}$ about  $\hat{\phi}$.

In this paper, we argue that there is a more intuitive approach that
better captures the nature of the problem. Indeed, in
Eq.~(\ref{eq:iso.rank.1}), $\vec{v}$ is a fixed vector while $R$ varies
over the entire space of solutions of Eqs.~(\ref{eq:so3.ind}). This implies
that we should view (\ref{eq:iso.rank.1}) \emph{not as an equation
  for $\vec{v}$, but rather as an equation for $R$}. We must ask
ourselves how the linear restriction (\ref{eq:iso.rank.1}) can be
satisfied by all solutions to (\ref{eq:so3.ind}) and, therefore, 
be no restriction at all. This is possible if and only if
(\ref{eq:iso.rank.1}) is trivial and, therefore, $\vec{v}=0$.

To be sure, we can demand that a non-zero vector
$\vec{u}\in\mathbb{R}^3$ be isotropic. However, the isotropy equation
$R\vec{u}=\vec{u}$ is not satisfied by all $R\in\text{SO}(3)$, since
$\vec{u}$ is clearly not invariant under a rotation about an axis
perpendicular to $\vec{u}$ itself. Therefore, in this case, the
isotropy equation for $\vec{u}$ must be satisfied together with
(\ref{eq:so3.ind}), restricting the valid solutions to the group
$\text{SO}(2)$ acting on the plane orthogonal to $\vec{u}$, which
constitutes a proper subgroup of $\text{SO}(3)$.

\paragraph{Ranks 2 and 3.}\label{sec:rank.2.3}

The existence of a rank-2 isotropic tensor $A$ gives rise to the
equality,
\begin{equation}
  \label{eq:iso.rank.2}
  R_{i_1j_1} R_{i_2j_2} A_{j_1j_2} = A_{i_1i_2},
  \qquad
  \forall R\in\text{SO}(3).
\end{equation}
Now, consider the system of equations for $R$ given by
(\ref{eq:so3.ind})(a), (\ref{eq:so3.ind})(b), and
(\ref{eq:iso.rank.2}). As anticipated in our alternative definition of
$\text{SO}(3)$ at the end of Section~2, the only way this system can
have the exact same set of solutions as Eqs.\ (\ref{eq:so3.ind}) alone
is if the quadratic equality (\ref{eq:iso.rank.2}) is not
independent of (\ref{eq:so3.ind})(a).  Therefore, we must have
$A_{ij}=\alpha \delta_{ij}$ for some $\alpha\in\mathbb{R}$. A
completely analogous argument shows that the rank-3 isotropic tensors
are of the form $A_{ijk}=\beta\varepsilon_{ijk}$ for some
$\beta\in\mathbb{R}$.

\paragraph{Rank 4.}\label{sec:rank.4}

We consider next a rank-4 isotropic tensor $A$, which leads to the
equality
\begin{equation}
  \label{eq:iso.rank.4}
  R_{i_1j_1} R_{i_2j_2} R_{i_3j_3} R_{i_4j_4} A_{j_1j_2j_3j_4} = A_{i_1i_2i_3i_4},
  \qquad
  \forall R\in\text{SO}(3).
\end{equation}
All solutions $R$ to the system of equations (\ref{eq:so3.ind}) will
identically satisfy a quartic relation like (\ref{eq:iso.rank.4}) if
and only if it is the product of (\ref{eq:so3.ind})(a) with
itself. Thus, in (\ref{eq:iso.rank.4}) the isotropic tensor must be of
the form $A_{j_1j_2j_3j_4}\sim\delta\delta$. However, we can contract
the indices in several different ways. We must group the four indices
$j_1,\ldots,j_4$ in (\ref{eq:iso.rank.4}) into an unordered set of two
unordered pairs, which can be accomplished in
$\frac{1}{2!}\binom{4}{2}\binom{2}{2}=3$ ways. Consequently, we
obtain,
\begin{equation}
  \label{eq:iso.rank.4.sol}
  A_{i_1i_2i_3i_4} = \alpha \delta_{i_1i_2}\delta_{i_3i_4} +
  \beta \delta_{i_1i_3}\delta_{i_2i_4} + \gamma
  \delta_{i_1i_4}\delta_{i_2i_3},
  \quad
  \alpha, \beta, \gamma \in \mathbb{R},
\end{equation}
which agrees with the well-known result in
\cite{jeffreys_1931,sant,hodge_1961,rmfe}.

We remark that Eq.\ (\ref{eq:iso.rank.4}) with $A$ equal to one
of the three tensors on the right-hand side of
(\ref{eq:iso.rank.4.sol}) corresponds directly to multiplying
(\ref{eq:so3.ind}) (a) by itself side-by-side with the indices
contracted in the three possible ways:
\[
  R_{i_1j}R_{i_2j}R_{i_3k}R_{i_4k} = \delta_{i_1i_2}\delta_{i_3i_4},
  \quad
  R_{i_1j}R_{i_3j}R_{i_2k}R_{i_4k} = \delta_{i_1i_3}\delta_{i_2i_4},
  \quad
  R_{i_1j}R_{i_4j}R_{i_2k}R_{i_3k} = \delta_{i_1i_4}\delta_{i_2i_3}.
\]

\paragraph{Rank 5.}\label{sec:rank.5}

For isotropic tensors of rank higher than four, index notation
becomes somewhat cumbersome. We write the equation expressing the
isotropy of a rank-$n$ tensor $A$ generically as
\begin{equation}
  \label{eq:iso.rank.n}
  R_{i_1j_1} R_{i_2j_2}\ldots R_{i_nj_n} A_{j_1j_2\ldots j_n} =
  A_{i_1i_2\ldots i_n},
 \qquad
  \forall R\in\text{SO}(3).
\end{equation}
We consider next rank-5 isotropic tensors. All solutions to
Eqs.~(\ref{eq:so3.ind}) automatically satisfy (\ref{eq:iso.rank.n})
with $n=5$ if it is the product of (\ref{eq:so3.ind})(a) and
(\ref{eq:so3.ind})(b). Therefore, in this case, the tensor in
(\ref{eq:iso.rank.n}) must be of the form
$A_{j_1j_2j_3j_4j_5}\sim\varepsilon\delta$. There are
$\binom{5}{2}\binom{3}{3}=10$ ways to partition five indices into an
unordered pair and an unordered trio; the internal ordering of the
latter is immaterial due to the antisymmetry of the Levi-Civita
tensor. Thus, we write
\begin{equation}
  \label{eq:iso.rank.5.sol}
  \begin{aligned}
  A_{i_1i_2i_3i_4i_5} & = \sum_I \alpha_I \varepsilon_{I_1} \delta_{I_2},
  \quad
  \alpha_I\in\mathbb{R},\\
  I = (I_1,I_2)&=
  (i_1i_2i_3,i_4i_5),(i_1i_2i_4,i_3i_5),(i_1i_2i_5,i_3i_4),(i_1i_3i_4,i_2i_5),(i_1i_3i_5,i_2i_4),\\
   &\quad\;\, (i_1i_4i_5,i_2i_3),(i_2i_3i_4,i_1i_5),(i_2i_3i_5,i_1i_4),(i_2i_4i_5,i_1i_3),(i_3i_4i_5,i_1i_2),
  \end{aligned}
\end{equation}
where for $I = (i_1i_2i_3,i_4i_5)$, for example, we have $I_1=i_1i_2i_3$ representing 
the indices of the Levi-Civita tensor and $I_2=i_4i_5$ representing the indices of 
the Kronecker $\delta$.

The results for rank-4 and rank-5 tensors are qualitatively
different. The set of three tensors in Eq.~(\ref{eq:iso.rank.4.sol}) is
linearly independent and therefore forms a basis for the space of rank-4
isotropic tensors. The ten tensors generated in Eq.~(\ref{eq:iso.rank.5.sol}),
on the other hand, constitute a linearly dependent spanning set rather than a
basis. This dependence arises from a certain algebraic relation
among the tensors in (\ref{eq:iso.rank.5.sol}), explicitly given by Eq.~(3.4) of
\cite{kea75}. From a practical standpoint, however, the most efficient way
to address this linear dependence is to introduce an inner
product on the space of rank-$n$ tensors,
\begin{equation}
  \label{eq:inner}
  \langle A,B \rangle = \sum_{i_1,i_2,\ldots, i_n=1}^{3}
  A_{i_1i_2\ldots i_n}^* B_{i_1i_2\ldots i_n}. 
\end{equation}
We can then compute the matrix of inner products,
$\langle A_i,A_j \rangle$, or Gram matrix, for the tensor sets appearing
in Eqs.~(\ref{eq:iso.rank.4.sol}) and~(\ref{eq:iso.rank.5.sol}), using 
Mathematica~\cite{wolf}. The rank of the Gram matrix is equal to the dimension 
of the subspace spanned by those tensors (see Theorem~7.2.10 in \cite{horn}). For the 
rank-4 case, we obtain a $3\times3$ non-singular matrix, confirming that those 
tensors form a linearly independent set. For the rank-5 case, however, the $10\times10$ 
Gram matrix is singular and has a rank of 6, as verified by the Mathematica command
\texttt{MatrixRank[]}. Therefore, only six of the tensors enumerated
in Eq.~(\ref{eq:iso.rank.5.sol}) can form a linearly independent
set. Restricting our choices to the first six tensors in
Eq.~(\ref{eq:iso.rank.5.sol}) yields a non-singular inner-product
matrix. These six rank-5 tensors agree exactly with the basis given in
Eq.~(4.2) of~\cite{kea75}.

\paragraph{Rank 6.}\label{sec:rank.6}

For ranks larger than 5, we only offer some general comments.  For
rank-6 isotropic tensors, two distinct algebraic structures arise: the
solutions to Eqs.~(\ref{eq:so3.ind}) automatically satisfy
(\ref{eq:iso.rank.n}) with $n=6$ if it is the product of
Eq.~(\ref{eq:so3.ind})(a) with itself three times, or the product
of Eq.~(\ref{eq:so3.ind})(b) with itself twice. Symbolically,
$A_{j_1\ldots j_6} \sim \delta\delta\delta +
\varepsilon\varepsilon$. We must take into account, however, the 
fundamental algebraic identity given by Eq.~(3.1) of~\cite{kea75},
which we repeat here for convenience:
\begin{equation}
  \label{eq:capelli.1}
  \varepsilon_{i_1i_2i_3}\varepsilon_{j_1j_2j_3} = 
  \begin{vmatrix}
    \delta_{i_1j_1} & \delta_{i_1j_2} & \delta_{i_1j_3} \\
    \delta_{i_2j_1} & \delta_{i_2j_2} & \delta_{i_2j_3} \\
    \delta_{i_3j_1} & \delta_{i_3j_2} & \delta_{i_3j_3}      
  \end{vmatrix}.
\end{equation}
This identity demonstrates that we need only consider
$A_{j_1\ldots j_6} \sim \delta\delta\delta$. Consequently, we must
partition the six indices $j_1,\ldots,j_6$ in (\ref{eq:iso.rank.n})
with $n=6$ into an unordered set of three pairs, which can be
accomplished in $\frac{1}{3!}\binom{6}{2}\binom{4}{2}\binom{2}{2}=15$
distinct ways. Evaluating these 15 rank-6 tensors via their Gram
matrix confirms that they form a linearly independent set. An explicit
basis consisting of these 15 tensors is detailed in Table~2
of~\cite{kea75}. For completeness, we note that although there are
$\frac{1}{2!}\binom{6}{3}\binom{3}{3}=10$ tensors of the form
$A_{j_1\ldots j_6} \sim \varepsilon\varepsilon$, their Gram matrix has
a rank of only 5. Thus, at most five of them can form a linearly
independent set, meaning any such choice would need to be supplemented
by ten additional tensors of the $\delta\delta\delta$ type to form a
complete basis.

\paragraph{Odd ranks. Rank 7.}\label{sec:rank.7}

In the case of tensors of odd rank $n=2k+1$, due to (\ref{eq:capelli.1}),
we only need to consider tensors of the form
$A_{i_1i_2\ldots i_n} \sim \delta\ldots\delta\varepsilon$,
involving the product of $k-1$ Kronecker deltas and one Levi-Civita
symbol. In this case, we group the $n$ indices into an unordered
set of $k-1$ unordered pairs and one unordered trio formed by the
remaining three indices. This can be accomplished in
\begin{equation}
  \label{eq:comb.1}
  \frac{1}{(k-1)!}\binom{2k+1}{2}\binom{2k-1}{2}\cdots\binom{5}{2}\binom{3}{3}
  = \frac{(2k+1)!}{(k-1)! 2^{k-1} 3!}
\end{equation}
distinct ways, which agrees exactly with Eq.~(2.3)
of~\cite{kea75}. For example, for rank-7 isotropic tensors, we observe
that all $\text{SO}(3)$ matrices satisfy Eq.~(\ref{eq:iso.rank.n})
identically with $n=7$ if it is the product of (\ref{eq:so3.ind})(a)
with itself and with (\ref{eq:so3.ind})(b). Thus, a spanning set for
rank-7 isotropic tensors can be constructed using all tensors of the
form $A\sim\delta\delta\varepsilon$. According to
Eq.~(\ref{eq:comb.1}), this yields 105 distinct tensors in the
spanning set for the space of rank-7 isotropic tensors. We see from
Table~1 of~\cite{kea75} that only 36 of these elements are linearly
independent. This result can be verified directly by computing the
rank of the $105\times105$ Gram matrix as discussed above. Such an
explicit basis is detailed in Table~3 of~\cite{kea75}.

\paragraph{Even ranks. Rank 8.}\label{sec:rank.8}

In the case of tensors of even rank $n=2k$, due to (\ref{eq:capelli.1}), 
we only need to consider tensors of the form 
$A_{i_1i_2\ldots i_n} \sim \delta\ldots\delta$, involving the product of 
$k$ Kronecker deltas. Thus, we partition the $n$ indices into an unordered 
set of $k$ unordered pairs, which can be accomplished in
\begin{equation}
  \label{eq:comb.2}
  \frac{1}{k!}\binom{2k}{2}\binom{2(k-1)}{2}\cdots\binom{2}{2}
  = \frac{(2k)!}{k! 2^{k}}
\end{equation}
distinct ways, which agrees exactly with Eq.~(2.2)
of~\cite{kea75}. For example, we observe that
Eq.~(\ref{eq:iso.rank.n}) with $n=8$ is satisfied identically if is
the product of (\ref{eq:so3.ind})(a) with itself four times, or the
product of (\ref{eq:so3.ind})(a) with (\ref{eq:so3.ind})(b)
twice. Symbolically, this means
$A_{j_1\ldots j_8} \sim \delta\delta\delta\delta +
\delta\varepsilon\varepsilon$.  As before, (\ref{eq:capelli.1})
implies that we need only consider the first of the two structures,
$A_{j_1\ldots j_8} \sim \delta\delta\delta\delta$;
Eq.~(\ref{eq:comb.2}) then yields exactly 105 distinct
tensors. However, the algebraic identity given by Eq.~(3.2)
of~\cite{kea75} implies that these tensors are not linearly
independent. The actual basis contains 91 tensors, which are detailed
explicitly in Table~5 of~\cite{kea75}.

\paragraph{Final remarks.}\label{sec:fin.rem}

In the foregoing, we pointed out that the isotropy Eqs.~(3) for the
Kronecker and Levi-Civita tensors constitute defining equations for
$\text{SO}(3)$. Therefore, any additional fundamental tensor would add
a further equation to that system, restricting the group to a proper 
subset of $\text{SO}(3)$. Similarly, we interpret the isotropy equations for
higher-rank tensors as equations for the rotation matrices themselves. In this
way, by bypassing the grueling finite rotations of
Jeffreys~\cite{jeffreys_1931} and the dense four-equation contraction
system of Hodge~\cite{hodge_1961}, we provide very
concise, pedagogically clear constructions of the spanning sets for
isotropic tensors of low ranks.

In the general case, a spanning set for isotropic tensors of even
rank $n=2k$ is formed by tensors $A\sim\delta\ldots\delta$ that represent the
tensor product of $k$ Kronecker deltas, while for odd rank $n=2k+1$ it is composed 
of tensors of the form $A\sim\delta\ldots\delta\varepsilon$ representing the tensor
product of $k-1$ Kronecker deltas and one Levi-Civita symbol. For
relatively low ranks, the index combinatorics can be worked out
explicitly by hand. Subsequently, the issue of linear dependence must be
addressed, either by means of the algebraic identities given in
\cite{kea75} or, alternatively, by the Gram-matrix method
advocated above as an efficient approach that lends itself well to
computerized symbolic or numeric evaluation.

Our discussion suggests possibly interesting research directions in connection with
isotropic tensors. On the one hand, for relatively low-rank tensors,
it would be highly desirable to develop an algorithm that yields linearly
independent bases directly, without the need to manage linearly
dependent spanning sets as a preliminary step. On the other hand, none of
the methods discussed above appear to be applicable to the case of
tensors of arbitrarily higher ranks, for which the number of spanning tensors
reaches $O(10^4)$ at ranks~11 and~12, and grow exponentially with the
rank. Certainly, these approaches are not viable for ranks in the
hundreds or thousands. One can naturally wonder what alternative methods would 
be effective in those regimes.

\paragraph*{Acknowledgements}

This work was partially supported by the Sistema Nacional de
Investigadores (SNI) of the Secretar\'ia de Ciencia, Humanidades,
Tecnolog\'ia e Innovaci\'on (SECIHTI) of M\'exico.



\begin{thebibliography}{99}
\setlength{\itemsep}{0pt}

\bibitem{jeffreys_1931}
H.\  Jeffreys, {\em Cartesian Tensors}, Cambridge University Press, Cambridge, 1931.

\bibitem{jeffreys_1973}
H.\  Jeffreys, On isotropic tensors, {\em Proc.\  Camb.\  Phil.\  Soc.} {\bf 73} (1973) 173.

\bibitem{sant} 
L.\  A.\  Santal\'o, {\em Vectores y tensores con sus aplicaciones}, EUDEBA (Editorial Universitaria de Buenos Aires), Buenos Aires, Argentina, 1961.

\bibitem{hodge_1961} 
P.\  G.\  Hodge, On isotropic cartesian tensors, {\em Am.\  Math.\  Mon.\ } {\bf 68} (1961) 793.

\bibitem{rmfe} 
O.\  Palillero-Sandoval, R.\  Carrada-Legaria, Y.\  E.\  Bravo-Garc\'ia, and E.\  Reynoso-Lara, Cartesian isotropic tensors for beginners, {\em Rev.\  Mex.\  Fis.\  E} {\bf 22} (2025) 020214.

\bibitem{kea75} E.\ A.\ Kearsley and J.\ T.\ Fong, Linearly
  independent sets of isotropic cartesian tensors of ranks up to
  eight, {\em J.\ Res.\ Nat.\ Bureau Standards} {\bf 79B} (1975) 49.

\bibitem{byron_1992} 
F.\  W.\  Byron and R.\  W.\  Fuller, \emph{Mathematics of Classical and Quantum Physics} (Dover Publications, New York, Rev.\  ed., 1992).
  
\bibitem{wolf} 
Wolfram Research Inc., {\em Mathematica}, Version 14.2, Champaign, IL, 2025.  

\bibitem{horn} 
R.\  A.\  Horn and C.\  R.\  Johnson, {\em Matrix Analysis}, Cambridge University Press, Cambridge, 1985.  

\end{thebibliography}
\end{document}